\documentclass[12pt]{article}
\usepackage{amsmath}
\usepackage{graphicx}
\usepackage{enumerate}
\usepackage{url} 

\usepackage{amssymb,verbatim}
\usepackage{epsfig}
\usepackage{fancyhdr}
\usepackage{pdflscape}
\usepackage{bm}
\usepackage{booktabs}
\usepackage{enumerate}
\usepackage{graphicx}
\usepackage{array}
\usepackage{amstext}
\usepackage{amsopn}
\usepackage{amsfonts}
\usepackage{amsmath}
\usepackage{mathrsfs}
\usepackage{amsbsy}
\usepackage{tablefootnote}
\usepackage{mathtools}
\usepackage{multicol}
\usepackage{multirow}
\usepackage{colortbl}
\usepackage{xcolor}
\usepackage{appendix}
\usepackage{caption}
\usepackage{algorithm,algpseudocode}
\usepackage{threeparttable}

\usepackage[authoryear,round]{natbib}
\usepackage[resetlabels]{multibib}
\usepackage{xr}
\newtheorem{theorem}{Theorem}
\newtheorem{lemma}{Lemma}
\newtheorem{corollary}{Corollary}
\newtheorem{remark}{Remark}

\newtheorem{cond}{Condition}
\newtheorem{model}{M}

\newcommand{\bx}{\boldsymbol{x}}
\newcommand{\bX}{\boldsymbol{X}}

\newcommand{\bW}{\boldsymbol{W}}

\newcommand{\bbR}{\mathbb{R}}
\newcommand{\bbE}{\mathbb{E}}
\newcommand{\bbN}{\mathbb{N}}

\newcommand{\calF}{\mathcal{F}}
\newcommand{\calJ}{\mathcal{J}}
\newcommand{\calL}{\mathcal{L}}

\newcommand{\calW}{\mathcal{W}}
\newcommand{\calX}{\mathcal{X}}
\newcommand{\calY}{\mathcal{Y}}

\newcommand{\btheta}{\bm{\theta}}
\newcommand{\bmu}{\bm{\mu}}
\newcommand{\bnu}{\bm{\nu}}
\newcommand{\bvartheta}{\bm{\vartheta}}

\def\argmax{\mathop{\rm argmax}}

\newcommand{\blind}{1}

\begin{document}

\def\spacingset#1{\renewcommand{\baselinestretch}%
{#1}\small\normalsize} \spacingset{1}


\if1\blind
{
  \title{\bf Deep Transformation Model}
  \author{Tong Wang *\\
    Department of Biostatistics, \\
    Yale School of Public Health, 
    New Haven, Connecticut, USA\\
    Shunqin Zhang \thanks{Tong Wang and Shunqin Zhang contributed equally to this work.} \\
    School of Mathematical Sciences, \\
    University of Chinese Academy of Sciences, China \\
    Sanguo Zhang\\
    Key Laboratory of Big Data Mining and Knowledge Management, \\
       Chinese Academy of Sciences, China \\
    Jian Huang\\
    Department of Applied Mathematics,\\ The Hong Kong Polytechnic University, Hong Kong SAR, China\\
    Shuangge Ma \\
    Department of Biostatistics, \\
    Yale School of Public Health, New Haven, Connecticut, USA}
  \maketitle
} \fi

\if0\blind
{
  \bigskip
  \bigskip
  \bigskip
  \begin{center}
    {\LARGE\bf Deep Transformation Model}
\end{center}
  \medskip
} \fi

\bigskip
\begin{abstract}
There has been a significant recent surge in deep neural network (DNN) techniques. Most of the existing DNN techniques have restricted model formats/assumptions. To overcome their limitations, we propose the nonparametric transformation model, which encompasses many popular models as special cases and hence is less sensitive to model mis-specification. This model also has the potential of accommodating heavy-tailed errors – a robustness property not broadly shared. Accordingly, a new loss function, which fundamentally differs from the existing ones, is developed. For computational feasibility, we further develop a double rectified linear unit (DReLU)-based estimator. To accommodate the scenario with a diverging number of input variables and/or noises, we propose variable selection based on group penalization. We further expand the scope to coherently accommodate censored survival data. The estimation and variable selection properties are rigorously established. Extensive numerical studies, including simulations and data analyses, establish the satisfactory practical utility of the proposed methods.
\end{abstract}

\noindent%
\vfill

\newpage
\spacingset{1.9} 

\section{Introduction}\label{sec:intro}
For accurately describing the relationship between a set of predictors $\bX$ and a response $Y$, a long array of statistical models and estimation and selection techniques have been developed. Some of them are robust (less sensitive) to model mis-specification, presence of outliers, etc. In the past decades, there has been a surge in deep neural network (DNN) techniques. Compared to ``classic’’ statistical models/estimations, DNNs can be advantageous under many scenarios with greater flexibility and superior prediction performance.

Multiple families of DNN models and estimations have been developed. Many of them have strong ties with regression models. For example, the squared loss has roots in linear regression and ordinary least squared estimation. The cross-entropy loss has roots in logistic regression and its maximum likelihood estimation. And the loss for censored survival data in \cite{zhong2022deep} is strongly connected to the partial likelihood of the Cox model. There are also DNN approaches, for example Generative Adversarial Networks (GANs), that target estimating distributions as opposed to specific model parameters. Despite of extensive developments, it has been recognized that new DNN models, loss functions, and estimation approaches are still critically needed. This study has been strongly motivated by the observation that, similar to regression analysis, DNN models are not immune from mis-specification, and estimation can be negatively impacted by outliers. There have been some recent efforts on robust DNNs such as the quantitle-based \citep{shen2021robust,Fan2023Factor}. However, to date, DNNs with robustness properties are still very limited. 

In practical data analysis, it is not uncommon to have multiple or a large number of predictors, some of which may be noises. Variable/model selection has been routinely conducted in regression analysis, which enhances interpretability, stability, cost-effectiveness, etc. DNNs are “black-box” with more complicated structures and more parameters and thus may have a stronger demand for variable/model selection. There have been a few efforts in the literature. For example, \cite{scardapane2017group} proposes applying group Lasso to either a specific network module or the entire network. \cite{lemhadri2021lassonet} develops LassoNet by adding a skip (residual) layer, which can control the participation of a predictor in any hidden layer. \cite{luo2023sparseinput} proposes a framework using group concave regularization for variable selection. 

Most of the existing DNN studies have focused on methodological, computational, and numerical developments. For a technique to be ``fully trustworthy’’, a strong theoretical foundation is needed. There have been a few recent efforts on developing the theoretical foundation of DNN techniques. For example, \cite{Bauer2019,Schmidt-Hieber2020,jiao2023AOS} establish the convergence rate of the DNN estimator based on the squared loss. \cite{Padilla2020QuantileRW,shen2024nonparametric} derive the convergence rate for quantile regression with the ReLU networks. \cite{shen2021robust} investigates the non-asymptotic upper bounds for regression models with heavy-tailed errors. And \cite{Farrell2021DNN} examines a general class of Lipschitz loss functions. 

This study advances from the existing DNN literature in multiple important perspectives. We propose a deep transformation model, which includes many popular models as special cases and can be more robust (less sensitive) to model mis-specification – this robustness is not commonly shared. For estimation, we propose a rank-based approach, which has roots in the maximum rank correlation (MRC) estimation and is well justified. It significantly differs from the existing DNN estimations and, as such, this study can fundamentally expand the framework of DNN estimation. As a ``byproduct’’, this estimation can accommodate heavy-tailed errors, possessing additional well-desired robustness. For computational feasibility, we propose a double rectified linear unit (DReLU)-based approximation. This novel approximation, compared to the alternatives, is more natural under the DNN framework and more efficient. To accommodate the scenario with a diverging number of predictors and/or noises, we apply penalization. Although penalization has been adopted in the literature, this study, with a new and more complicated model and estimation, can significantly expand its utilization. Unlike many others, the proposed model/estimation can coherently accommodate right censored response under the same framework. We rigorously establish important properties including identifiability and estimation and selection consistency. Beyond providing a uniquely strong basis to the proposed model and estimates, this effort can also solidify the theoretical foundation of DNNs. Last but not least, as shown in the numerical examples, this study can deliver a practically useful new DNN tool that can have broad applications.

\section{Methods}\label{sec:model}
\subsection{Nonparametric Transformation Model}
For $(\bX,Y)\in \calX\times\mathcal{Y}\subset \bbR^{p}\times \bbR$, where $\bX $ is the $p$-vector of predictors and $Y $ is the response variable, we consider the nonparametric transformation model:
\begin{align}\label{Model-proposed}
    Y=D\circ F(f^{\ast}(\bX),\epsilon),
\end{align}
where $D:\bbR\to\calY$ is a non-degenerate increasing function, $F:\bbR^2\to\bbR$ is strictly increasing regarding each of its coordinate, $f^{\ast}:\calX\to\bbR$ is the target function, and $\epsilon$ is a random variable independent of $\bX$. 

This model and corresponding developments have been motivated by the observation that, despite of greater flexibility, DNNs are not immune from model mis-specification. The proposed model accommodates multiple types of response including continuous, categorical, and survival under a single umbrella. It includes many models -- including but not limited to linear, generalized linear, single-index, multi-index, nonparametric regression, additive regression, Cox, and others – as special cases. As such, it is comparatively much less sensitive to model mis-specification.

\subsection{Deep Rank Estimation}\label{subsec:nn-mrc}
Let $\{(\bX_i,Y_i)\}_{i=1}^n$ denote $n$ independent and identically distributed (i.i.d.) random copies of $(\bX, Y)$. Consider the rank-based objective function: 
\begin{align}\label{obj:NMRC}
    U_{n}(f)=\frac{1}{n(n-1)}\sum_{i\neq j}I(Y_i>Y_j)I(f(\bX_i)>f(\bX_j)),
\end{align}
where $I(\cdot)$ is the indicator function. Without loss of generality, we assume that $\|f\|_{\infty}=1$, where $\|f\|_{\infty}\coloneqq\sup_{\bx\in\calX}|f(\bx)|$. 

We propose estimation using a Feedforward Neural Network (FNN), denoted as $f_{\btheta}$, where $\btheta$ represents the network parameters (weights and biases). The detailed architecture is elucidated in Section \ref{subsec:imp}. Consequently, the proposed estimator is defined as:
\begin{align}\label{est:NMRC}
    \hat{f}_{\btheta}\in\argmax_{f_{\btheta}\in\calF_{\btheta}} U_n(f_{\btheta}),
\end{align}
where $\calF_{\btheta}$ is a function class consisting of FNNs. 

The objective function (\ref{obj:NMRC}) measures the rank correlation between $Y$ and $f(\bX)$. In regression, the maximum rank correlation (MRC) estimation has been well examined. In most of such studies \citep{han2015provable,fan2020rank,shen2023linearized}, $f(\bX)$ takes a linear form. There are also a handful of studies that consider more complicated forms \citep{Horowitz2001,nguelifack2020robust}, for example, nonparametric additive. The DNN-based estimation is the natural next step. However, with a much different and more challenging framework, it significantly advances from the existing literature.

\subsection{Deep Double ReLU  Approximation}\label{subsec:smooth-nn-mrc}

For computational feasibility, in the existing studies, the Sigmoid and other approximations \citep{Song2006,lin2013smoothed} have been adopted to approximate the indicator function. The DNN estimation significantly differs from regression. We propose a new approximation, which, under the DNN framework, has computational and theoretical advantages. 

\begin{figure}[t]
    \centering
    \includegraphics[width=0.21\textheight]{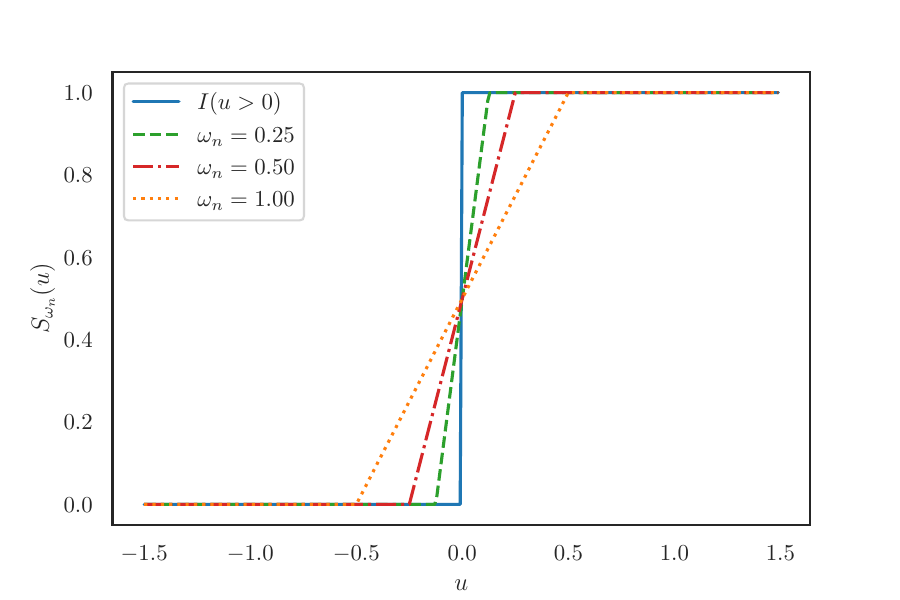}
    \includegraphics[width=0.21\textheight]{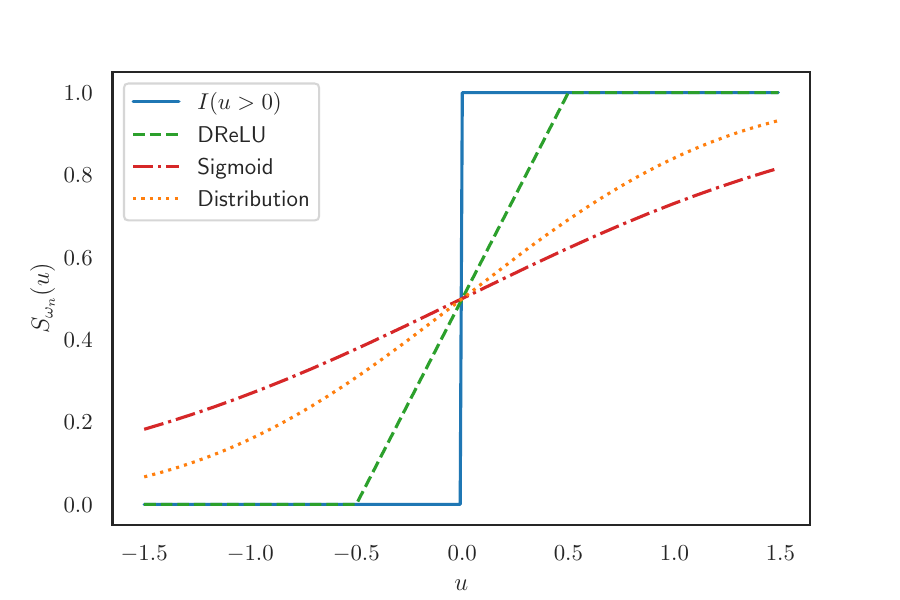}
    \includegraphics[width=0.21\textheight]{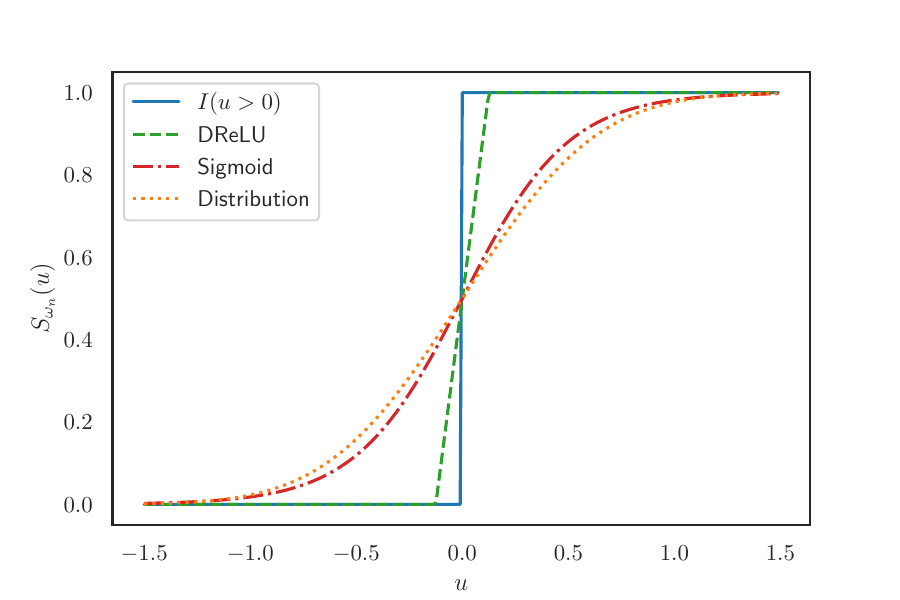}
    \caption{Approximation of the indicator function. Left: DReLU with different $\omega_n$ values. Middle: approximations with $\omega_n=1.00$. Right: approximations with $\omega_n=0.25$.}
    \label{fig:smooth}
\end{figure}

Consider the Double Rectified Linear Unit (DReLU) function: 
\begin{align}\label{eq:S-omega}
    S_{\omega_n}(u)=\sigma(u/\omega_n+1/2)-\sigma(u/\omega_n-1/2),\  u\in\bbR,
\end{align}
where $\sigma(\cdot)=\max(\cdot,0)$ is the rectified linear unit (ReLU) activation function, and $\omega_n$ is a small positive tuning parameter. This approximation is visually presented in Figure \ref{fig:smooth}, along with the alternatives adopted in the literature.

Accordingly, we propose the approximated objective function and estimate:
\begin{align}\label{obj:SNMRC}
    U_{n,\omega_n}(f)=&\frac{1}{n(n-1)}\sum_{i\neq j}I(Y_i>Y_j)S_{\omega_n}(f(\bX_i)-f(\bX_j)),
\end{align}
\begin{align}\label{est:SNMRC}    \hat{f}_{\btheta_{\omega_n}}\in\argmax_{f_{\btheta}\in\calF_{\btheta}}U_{n,\omega_n}(f_{\btheta}).
\end{align}

\begin{remark}
ReLU is a foundational component of DNN. With its computational advantages, DReLU circumvents the vanishing gradients problem, thereby mitigating optimization challenges and providing a stable computational basis. By employing simpler linear thresholds and eliminating expensive operations, this approximation has a lower computational cost than some others such as the Sigmoid and normal cumulative distribution function approximations. As shown below, it also has competitive theoretical properties. 
\end{remark}

\subsection{Implementation}\label{subsec:imp}
We conduct estimation using FNN denoted as $f_{\btheta}:\bbR^{p}\to\bbR$. This network is formulated as a composition of a series of functions: 
\begin{align}\label{eq:FNN}
    f_{\boldsymbol{\theta}}(\bx)=\bW_{\calL}\sigma(\bW_{\calL-1}\sigma(\ldots\bW_1\sigma(\bW_0\bx+\bm{b}_0)+\bm{b}_1)+\ldots)+\bm{b}_{\calL-1})+\bm{b}_{\calL}, \ \bx \in \mathbb{R}^{p_0},
\end{align}
where $\sigma(\cdot)$ is the ReLU activation function, and $\calL$ is the depth of the network. 
For $i\in\{0,1,\ldots,\calL\}$, $\bW_i \in \bbR^{p_{i+1}\times p_{i}}$ and $\bm{b}_i \in \mathbb{R}^{p_{i}}$ are the weight matrix and bias vector of the $i$-th layer, respectively, and $p_i$ is the width (the number of neurons or computational units) of the $i$-th layer. 
Define the network parameters  as $\btheta=\{(\bW_i,\bm{b}_i):i=0,1,\ldots,\calL\}$. 
The input dimension is $p_0 = p$, and the output dimension is $p_{\calL+1}=1$. 

Denote the data by $\mathcal{D}\coloneqq \{(\bx_i,y_i)\}_{i=1}^n$. 
Let $\{(\bx^{(l)}_i,y^{(l)}_i)\}_{i=1}^m$ be the set of samples randomly selected for the $l$-th batch, where the batch size is $m$. Thus, each epoch comprises $\lceil n/m\rceil$ batches, with $\lceil\cdot\rceil$ representing the ceiling function. The neural network, updated after the $k$-th epoch, is denoted by $f_{\btheta^{(k)}}$. The essential training steps are presented in Algorithm \ref{alg1}. To enhance training efficiency and model performance, and to prevent overfitting, several widely utilized DNN training strategies are implemented, including weight decay, early stopping, learning rate decay, and random dropout. The selection of hyperparameters and neural network architectures varies across data and is described in Section A.1.

\begin{algorithm}
	\caption{Deep DReLU Rank Estimation}
	\label{alg1}
\begin{algorithmic}[1]
\State {\bfseries Require:} Data $\mathcal{D}$; Hyperparameter $\omega_n$; Batch size $m$; Training epoch $K$.

\State Initialize $f_{\btheta^{(0)}}$, and compute $f_{\btheta^{(0)}}(\boldsymbol{x}_1),\ldots,f_{\btheta^{(0)}}(\boldsymbol{x}_n)$
\For {$k=1$ \textbf{to} $K$}
\For {$l=1$ \textbf{to} $\lceil n/m \rceil$}
\State Sample $\{(\bx^{(l)}_i,y^{(l)}_i)\}_{i=1}^m$ from data $\mathcal{D}$ without replacement.
\State Update $f_{\btheta}$ by ascending its stochastic gradient: 
\begin{align*}
    \nabla_{\btheta}\frac{1}{2m(n-1)} \sum_{i=1}^m \sum_{j=1}^n S_{\omega_n}\left(\operatorname{sgn}\left(y_i^{(l)}-y_j\right)\left(f_{\btheta}\left(\bx_i^{(l)}\right)-f_{\btheta^{(k-1)}}\left(\bx_j\right)\right)\right),
\end{align*}
where $\operatorname{sgn}(\cdot)$ is the sign function.
\EndFor
\State Compute $f_{\btheta^{(k)}}(\boldsymbol{x}_1),\ldots,f_{\btheta^{(k)}}(\boldsymbol{x}_n)$
\EndFor
\end{algorithmic}
\end{algorithm}

\section{Statistical Properties}\label{sec:theorey}

\subsection{Identifiability}\label{subsec:ident}
The population-level objective functions for the rank estimator and the DReLU estimator are respectively defined as:
\begin{align}
    U(f)=&\bbE[I(Y_1>Y_2)I(f(\bX_1)>f(\bX_2))],\label{obj:pop-NMRC}\\
    U_{\omega_n}(f)=&\bbE[I(Y_1>Y_2)S_{\omega_n}(f(\bX_1)-f(\bX_2))].\label{obj:pop-S-NMRC}
\end{align}
The following conditions are assumed.

\begin{cond}\label{C:mathcal-F}
    The target function $f^{\ast}$ belongs to the class $\mathcal{F}=\{f:\calX \to\bbR, \|f\|_{\infty}=1\}$ that satisfies: (i) any $f\in\calF$ is Lipschitz continuous and strictly increasing in the first coordinate of $\bX$; (ii) $\calF$ is compact with respect to the metric $d(f_1,f_2)=\|f_1-f_2\|_{\infty}$; 
    (iii) no two functions in $\calF$ are strictly increasing transformations of each other.
\end{cond}

\begin{cond}\label{C:support X}
    The conditional probability of the first coordinate of $\bX$ has a Lebesgue density that is everywhere positive over $\calX$, conditional on the other coordinates of $\bX$.
\end{cond}

\begin{cond}\label{C:strictly-mono}
    For any $\bX_i,\bX_j\in\calX$ with $f^{\ast}(\bX_i)>f^{\ast}(\bX_j)$, there exists a constant $c_0\in\mathbb{R}$ such that $\Pr(D\circ F(f^{\ast}(\bX_i),\epsilon)\geq c_0| \bX_i,\bX_j)>\Pr(D\circ F(f^{\ast}(\bX_j),\epsilon)\geq c_0| \bX_i,\bX_j),$  where the probability is taken with respect to $\epsilon$.
\end{cond}

Condition \ref{C:mathcal-F} (i)-(ii) and Condition \ref{C:support X} are substitutions of the parametric space restrictions in \cite{han1987non}.  Condition \ref{C:mathcal-F} (iii) is satisfied by many commonly used functions, such as linear, homogeneous, norm, and polynomial functions. The class of FNNs without the bias vector also satisfies Condition \ref{C:mathcal-F} (iii). Condition \ref{C:strictly-mono} ensures that variations in $f^{\ast}$ correspond to variations in the conditional distribution of $Y$ given $\bX$. This condition is met if $\epsilon$ does not follow a singular distribution. The same condition is required in \cite{matzkin2007nonparametric}.

\begin{lemma}\label{lem:ident-U}
Under Conditions \ref{C:mathcal-F}-\ref{C:strictly-mono}, the target function $f^{\ast}$ is the unique maximizer of $U(f)$.
\end{lemma}

\begin{lemma}\label{lem:ident-U-omega}
Suppose that Conditions \ref{C:mathcal-F}-\ref{C:strictly-mono} hold. Given $\omega_n>0$, there exists a function $f^{\ast}_{\omega_n}$ that attains the maximum of $U_{\omega_n}(f)$ uniquely over $\calF$, and  $d(f^{\ast}_{\omega_n},f^{\ast})\to0$ as $\omega_n\to0$.
\end{lemma}

These two lemmas establish identifiability without and with the approximation. Lemma \ref{lem:ident-U-omega} can be extended to other approximations such as Sigmoid, with slightly more stringent conditions. Proof is provided in Section \ref{subsec:ident} in supplementary materials.

\subsection{Convergence properties with a fixed $p$}\label{subsec:convergence-low}
Assuming $p$ is fixed, we examine the convergence rates of excess risk for both the deep rank estimator $\hat{f}_{\btheta}$ and the deep DReLU estimator $\hat{f}_{\btheta_{\omega_n}}$. Additional conditions are needed.

\begin{cond}\label{C:X-support-bound}
    The support of $\bX$ is a bounded compact set in $\mathbb{R}^{p}$. Without loss of generality, assume that $\mathcal{X}=[0,1]^p$.
\end{cond}

\begin{cond}\label{C:holder}
    The function class $\calF$ belongs to the H\"older class $\mathcal{H}^{\beta}(\calX,B_0)$ for a given $\beta>0$ and a finite constant $B_0>0$ defined as:
        \begin{align*}
        \mathcal{H}^{\beta}(\calX,B_0)=\left\{f:\calX\to\bbR,\max_{\|\bm{\pi}\|_1\leq \lfloor\beta\rfloor}\|\partial^{\bm{\pi}}f\|_{\infty}\leq B_0,\max_{\|\bm{\pi}\|_1=\lfloor\beta\rfloor}\sup_{\bm{r},\bm{s}\in\calX,\bm{r}\neq \bm{s}}\frac{|\partial^{\bm{\pi}}f(\bm{r})-\partial^{\bm{\pi}}f(\bm{s})|}{\|\bm{r}-\bm{s}\|^{\beta-\lfloor\beta\rfloor}_{2}}\leq B_0\right\},
    \end{align*}
    where $\partial^{\bm{\pi}}=\partial^{\pi_1}\ldots\partial^{\pi_p}$ with $\bm{\pi}=(\pi_1,\ldots,\pi_p)^{\top}\in\mathbb{N}^{p}_0$, $\|\bm{\pi}\|_1=\sum_{i=1}^{p}\pi_i$,  $\mathbb{N}_0$ is the set of non-negative integers, and $\lfloor\cdot\rfloor$ is the floor function.
\end{cond}

\begin{cond}\label{C:indicator-bound}
    For any function $\delta(\cdot):\mathcal{X}\to\mathbb{R}$,  the target function $f^{\ast}$ satisfies that, for any paired samples $(\bX_i,Y_i)$ and $(\bX_j,Y_j)$,
    \begin{align*}
        \bbE_{Y}[\Pr(f^{\ast}(\bX_i)> f^{\ast}(\bX_j)|Y_i,Y_j)-\Pr(f^{\ast}(\bX_i)+\delta(\bX_i)> f^{\ast}(\bX_j)+\delta(\bX_j)|Y_i,Y_j)]\leq c_1\|\delta\|_{\infty},
    \end{align*}
    where $c_1$ is a positive constant.
\end{cond}

Condition \ref{C:X-support-bound} requires the bounded support of $\bX$. Condition \ref{C:holder}  prescribes the smoothness of $f^{\ast}$. They have been commonly assumed in nonparametric regression including DNN studies   \citep{Bauer2019, jiao2023AOS}. Condition \ref{C:indicator-bound} addresses the discontinuity of the indicator function and is not needed for the DReLU estimator. Similar conditions have been assumed in nonparametric studies such as \cite{Padilla2020QuantileRW}.


The network function class $\calF_{\btheta}$, which consists of FNNs $f_{\btheta}$ defined in (\ref{eq:FNN}), is characterized by parameter $\btheta$, depth $\calL$, width $\mathcal{W}$ (maximum width of hidden layers), size $\mathcal{S}$ (total number of parameters), and bound $\mathcal{B}$.  We further define $\calL$ and $\calW$ as:
\begin{align}\label{eq:depth-width-NN-low-p}
    \mathcal{L}=21 (\lfloor\beta\rfloor+1)^2 M\lceil\log_2 (8M)\rceil +2p, \ \mathcal{W}=38 (\lfloor\beta\rfloor+1)^2 3^p p^{\lfloor\beta\rfloor+1}N\lceil\log_2 (8N)\rceil,
\end{align}
for any $M,N\in\mathbb{N}_{+}$, where $\mathbb{N}_{+}$ is the set of positive integers. 
The convergence rate results are established as follows.

\begin{theorem}\label{thm:convergence-rate-low}
     Suppose that Conditions \ref{C:mathcal-F}-\ref{C:indicator-bound} hold and  $\calF_{\btheta}$ satisfies (\ref{eq:depth-width-NN-low-p}). Then, 
     \begin{align}
         \bbE[U(f^{\ast})-U(\hat{f}_{\btheta})]\precsim & (N M)^{-2 \beta / p}+(N\log N) (M\log M) \sqrt{\frac{\log(NM)\log n}{n}},\label{eq:excess-thm1-state}
     \end{align}
     where ``$\precsim$" indicates that the left side is upper bounded by the right side up to a multiplicative constant,  and $\mathbb{E}$ is taken with respect to $\{(\bX_i,Y_i)\}_{i=1}^{n}$. 
     Further, if we set $MN=O(n^{\frac{p}{2 (p+2\beta)}})$, (\ref{eq:excess-thm1-state}) implies that 
 \begin{align}\label{eq:excess-low}
        \bbE[U (f^{\ast})-U(\hat{f}_{\btheta})]\precsim n^{-\frac{\beta}{2\beta+p}}\log^3 n.
 \end{align}
\end{theorem}


\begin{theorem}\label{thm:convergence-rate-low-smooth-slower}
    Suppose that Conditions \ref{C:mathcal-F}-\ref{C:holder} hold and  $\calF_{\btheta}$ satisfies (\ref{eq:depth-width-NN-low-p}). Then, 
   \begin{align}\label{eq:thm-CR-S-NoPen-state-1}
    \nonumber & \bbE[U(f^{\ast})-U(\hat{f}_{\btheta_{\omega_n}})]\\
    \precsim & \omega_n + \omega_n^{-1}(NM)^{-2\beta/p}+(N\log N)(M \log M)\sqrt{\frac{\log(NM)(\log n-\log \omega_n)}{n}}.
    \end{align}
    Further, if we set $MN=O(n^{\frac{p}{2\beta+2p}})$ and $\omega_n=O(n^{-\frac{\beta}{2\beta+2p}})$, (\ref{eq:thm-CR-S-NoPen-state-1}) implies that
    \begin{align}\label{eq:thm-CR-S-NoPen-state-2}
         \mathbb{E}[U(f^{\ast})-U(\hat{f}_{\btheta_{\omega_n}})]\precsim n^{-\frac{\beta}{2\beta+2p}}\log^3 n.
    \end{align}
\end{theorem}

Theorem \ref{thm:convergence-rate-low-smooth-slower} suggests that, compared to $\hat{f}_{\btheta}$, the results for $\hat{f}_{\btheta_{\omega_n}}$ can be established under milder conditions, albeit with a slight loss in convergence rate. The approximation of the indicator function introduces an additional term to the upper bound of the excess risk, denoted by $\mathcal{E}_{\omega_n}$, which equals $\omega_n$ in (\ref{eq:thm-CR-S-NoPen-state-1}). The convergence rate is influenced by the trade-off between $\mathcal{E}_{\omega_n}$ and the neural network approximation error. As described in Corollary 1 in supplementary materials, the same rate can be achieved under equivalent conditions.  In Table \ref{tab:smooth-error}, we compare the DReLU approximation with two alternatives, whose results are established in Corollaries 2-4 in supplementary materials.  

\begin{table}[t]
    \centering
    \caption{Error $\mathcal{E}_{\omega_n}$ and excess risk under different approximations.}\label{tab:smooth-error}
    \begin{threeparttable}
    \resizebox{\textwidth}{!}{\begin{minipage}{\textwidth}
    \begin{tabular}{ccccc}
    \hline
   Approximation & $S_{\omega_n}(u)$ & 
$\mathcal{E}_{\omega_n}$ & $\mathcal{E}_{\omega_n}/\omega_n$ &  $\mathbb{E}[U(f^{\ast})-U(\hat{f}_{\btheta_{\omega_n}})]$ \\
   \hline
DReLU & $\sigma(\frac{u}{\omega_n}+\frac{1}{2})-\sigma(\frac{u}{\omega_n}-\frac{1}{2})$  & $\omega_n$ & 1 & $n^{-\frac{\beta}{2\beta+2p}}\log^3 n$ \\
   Sigomid & $1/\{1+\exp(-\frac{u}{\omega_n})\}$ &$\exp(-\omega^{a-1}_n)+\omega_n^{a}$ &  $\to \infty$ &$n^{-\frac{ a\beta}{2\beta+2p}}\log^3 n$ \\
    Distribution &$\frac{1}{\sqrt{2\pi}}\int_{-\infty}^{{u}/{\omega_n}}e^{-t^2/2}dt$   & $\int_{-\infty}^{-\omega^{a-1}_n}e^{-t^2}dt+\omega_n^{a}$ &  $\to \infty$ & $n^{-\frac{a\beta}{2\beta+2p}}\log^3 n$\\
    \hline
    \end{tabular}  
    \footnotesize
    Notes: $0<a<1$ is a nuisance parameter. ``$\to\infty$" in the fourth column indicates convergence to infinity as $\omega_n\to0$.
    The last column shows the upper bound of excess risk when $\omega_n=O(n^{-\frac{\beta}{2\beta+2p}})$. 
    \end{minipage}}
    \end{threeparttable}
\end{table}

\subsection{Convergence properties with a diverging $p$}\label{subsec:convergence-high}

When $p$ diverges with $n$,  sparsity is often assumed. Denote the set of important predictors as $\bX_{\calJ}$ with $\calJ\subset \{1,\ldots,p\}$ and its cardinality as $|\calJ|\coloneqq p_s$. As in the literature \citep{ feng2017sparse,Fan2023Factor}, we consider the scenario with a fixed $p_s$. To simplify notation, we assume that $\calJ=\{1,\ldots,p_s\}$. The set of unimportant predictors is denoted as $\bX_{\calJ^{c}}$ and its cardinality as $|\calJ^{c}|\coloneqq p_c$. Formally, the following sparsity condition is assumed.

\begin{cond}\label{C:sparsity}
For the target function $f^{\ast}:\calX\to\bbR$, there exists function  $\tilde{f}^{\ast}:\calX_{\calJ}\to\bbR$ belonging to the H\"older class $\mathcal{H}^{\beta_1}(\calX_{\calJ},B_1)$ with   $\beta_1>0$ and $0<B_1<\infty$ such that 
    \begin{align}\label{eq:low-complexity-form}
    f^{\ast}(\bX)=\tilde{f}^{\ast}(X_1,\ldots,X_{p_s}),
    \end{align}
    where $\calX_{\calJ}$ is the support of $\bX_{\calJ}$.
\end{cond}

Similar conditions have been assumed \citep{Schmidt-Hieber2020,Fan2023Factor,wu2023neural}. For depth $\calL$ and width $\calW$ of network class $\calF_{\btheta}$, we specify:
\begin{align}\label{eq:depth-width-NN-high-p}
    \calL = 21 (\lfloor\beta_1\rfloor+1)^2 M\lceil\log_2 (8M)\rceil +2p_s, \ \mathcal{W}=38p(\lfloor\beta_1\rfloor+1)^2 3^{p_s} {p_s}^{\lfloor\beta_1\rfloor+1}N\lceil\log_2 (8N)\rceil,
\end{align}
for any $M,N\in\bbN_{+}$.  The convergence results are established as follows.

\begin{theorem}\label{thm:convergence-rate-high}
      Suppose that Conditions \ref{C:mathcal-F}-\ref{C:sparsity} hold and  $\calF_{\btheta}$ satisfies (\ref{eq:depth-width-NN-high-p}). Then, 
     \begin{align}\label{eq:thm2-rate}
    \bbE[U(f^{\ast})-U(\hat{f}_{\btheta})] \precsim (N M)^{-2 \beta_1 / p_s} + p(N\log N) (M\log M) \sqrt{\frac{\log(pNM)\log n}{n}},
\end{align}
which converges to zero under suitable choices of $N$ and $M$, provided that $p\ll n^{1/2}$.
\end{theorem}

\begin{theorem}\label{thm:convergence-rate-high-smooth}
      Suppose that Conditions \ref{C:mathcal-F}-\ref{C:holder} and \ref{C:sparsity} hold, and  $\calF_{\btheta}$ satisfies (\ref{eq:depth-width-NN-high-p}). Then,
     \begin{align}\label{eq:thm2-rate-smooth}
    \nonumber&\bbE[U(f^{\ast})-U(\hat{f}_{\btheta_{\omega_n}})] \\
    \precsim &\omega_n + \omega_n^{-1}(N M)^{-2 \beta_1 / p_s} + p(N\log N) (M\log M) \sqrt{\frac{\log(pNM)(\log n-\log \omega_n)}{n}},
\end{align}
which converges to zero under suitable choices of $N$, $M$ and $\omega_n$, provided that $p\ll n^{1/2}$.
\end{theorem}

\begin{table}[t]
    \centering
    \caption{Convergence rates of excess risk for the proposed estimators with various $p$.}
    \label{tab:convergence-high}
    \begin{threeparttable}
    \resizebox{\textwidth}{!}{\begin{minipage}{\textwidth}
    \begin{tabular}{c ccc c ccc}
    \hline
    & \multicolumn{3}{c}{$\hat{f}_{\btheta}$} & & \multicolumn{3}{c}{$\hat{f}_{\btheta_{\omega_n}}$}\\
    \cline{2-4}\cline{6-8}
    $p$& $MN$ & $\omega_n$ & $\bbE[U(f^{\ast})-U(\hat{f}_{\btheta})]$& & $MN$ & $\omega_n$ & $\bbE[U(f^{\ast})-U(\hat{f}_{\btheta_{\omega_n}})]$\\
    \hline
    $O(1)$ & $n^{\frac{p_s}{2(2\beta_1+p_s)}}$ & - & $n^{-\frac{\beta_1}{2\beta_1+p_s}} \log^3 n $ && $n^{\frac{p_s}{2\beta_1+2p_s}}$ & $n^{-\frac{\beta_1 }{2\beta_1+2p_s}}$ & $n^{-\frac{\beta_1}{2\beta_1+2p_s}} \log^3 n$\\
    $\log^{a_1}n$ & $n^{\frac{p_s}{2(2\beta_1+p_s)}}$ & - & $n^{-\frac{\beta_1}{2\beta_1+p_s}} \log^{a_1+3} n$ && $n^{\frac{p_s}{2\beta_1+2p_s}}$ & $n^{-\frac{\beta_1}{2\beta_1+2p_s}}$ & $n^{-\frac{\beta_1}{2\beta_1+2p_s}} \log^{3+a_1} n$\\
    $n^{a_2}$ & $n^{\frac{p_s}{2\beta_1+p_s}(\frac{1}{2}-a_2)}$ & - & $n^{-\frac{\beta_1}{2\beta_1+p_s}(\frac{1}{2}-a_2)} \log^3 n$&& $n^{\frac{p_s}{\beta_1+p_s}(\frac{1}{2}-a_2)}$ & $n^{-\frac{\beta_1}{\beta_1+p_s}(\frac{1}{2}-a_2)}$ & $n^{-\frac{\beta_1}{\beta_1+p_s}(\frac{1}{2}-a_2)} \log^3 n$\\
    \hline     
    \end{tabular}
    \footnotesize
    Notes: $a_1$ and $a_2$ are positive constant satisfying $0<a_1<\infty$ and $0<a_2<1/2$.
    \end{minipage}}
    \end{threeparttable}
\end{table}

In Table \ref{tab:convergence-high}, we present a few special cases of the general results obtained in Theorems \ref{thm:convergence-rate-high}-\ref{thm:convergence-rate-high-smooth}. Convergence slows down as $p $ increases but can be achieved as long as $p\ll n^{1/2}$.  Similar to the fixed $p$ case, compared to the rank estimator, the DReLU estimator can achieve a slower convergence rate under milder conditions and the same rate under the same conditions. The convergence properties with the Sigmoid and distribution function approximations are established in Corollary 5 and Table A.1 in supplementary materials.

\section{Penalized Variable Selection}\label{sec:variable-selection}

\subsection{Methods}
For variable selection, we propose the penalized deep rank objective function and the penalized DReLU objective function: 
\begin{align}
    U_{n,\lambda_n}(f_{\btheta})=&\frac{1}{n(n-1)}\sum_{i\neq j}I(Y_i>Y_j)I(f_{\btheta}(\bX_i)>f_{\btheta}(\bX_j))-\lambda_n \sum_{q=1}^{p}\|\bW_{0,q}(\btheta)\|_2,\label{obj:P-NMRC}\\
    U_{n,\omega_n,\lambda_n}(f_{\btheta})=&\frac{1}{n(n-1)}\sum_{i\neq j}I(Y_i>Y_j)S_{\omega_n}(f_{\btheta}(\bX_i)-f_{\btheta}(\bX_j))-\lambda_n \sum_{q=1}^{p}\|\bW_{0,q}(\btheta)\|_2,\label{obj:P-NMRC-smooth}
\end{align}
where $\lambda_n$ is a data-dependent tuning parameter, $\bW_0 (\btheta)$ is the first-layer weight matrix of network $f_{\btheta}$, and $\bW_{0,q}(\btheta)$ represents the $q$-th column of $\bW_0 (\btheta)$, linking the $q$-th component of $\bX$. The corresponding penalized estimators are defined as: 
\begin{align}\label{est:P-NMRC}
    \hat{f}_{\btheta_{\lambda_n}}\in\argmax_{f_{\btheta}\in\calF_{\btheta}} U_{n,\lambda_n}(f_{\btheta}),\text{ and }\hat{f}_{\btheta_{\omega_n,\lambda_n}}\in\argmax_{f_{\btheta}\in\calF_{\btheta}} U_{n,\omega_n,\lambda_n}(f_{\btheta}).
\end{align}

The network parameters of $\hat{f}_{\btheta_{\lambda_n}}$ and $\hat{f}_{\btheta_{\omega_n,\lambda_n}}$ are denoted as $\hat{\btheta}_{\lambda_n}$ and $\hat{\btheta}_{\omega_n,\lambda_n}$, respectively, with 
their first-layer weights denoted by $\bW_0 (\hat{\btheta}_{\lambda_n})$ and $\bW_0 (\hat{\btheta}_{\omega_n,\lambda_n})$. 
The components of $\bX$ with nonzero first-layer weights are identified as important \citep{feng2017sparse}.
Here, the first-layer weights corresponding to a specific predictor are viewed as a group. Selection is achieved with the all-in-or-all-out property of group Lasso.

\subsection{Statistical properties}\label{subsec:thm-variable-selection}
To establish selection results, beyond the convergence rates of excess risk, we also need to examine the convergence rates at the parameter level. Define:
\begin{align}\label{eq:Theta-def}
    \Theta\coloneqq \{\btheta:U(f_{\btheta})=U(f^{\ast}_{\btheta})\},
\end{align}
where $f^{\ast}_{\btheta}\in\argmax_{f_{\btheta}\in\calF_{\btheta}}U(f_{\btheta})$,  
and $U(\cdot)$ is the population-level objective function defined in (\ref{obj:pop-NMRC}).
The set $\Theta$ encompasses the network parameters $\btheta$ such that the networks parameterized by them achieve the same maximum of $U(\cdot)$ as $f^{\ast}_{\btheta}$. 
Under appropriate conditions, $f^{\ast}_{\btheta}$ can closely approximate the target function $f^{\ast}$. Due to the inherent unidentifiability of network parameters, there exists a subset within $\Theta$ that distinguishes the important variables via the first-layer weights.
Moreover, for any $\btheta$ within $\Theta$, the relationship $\btheta\in\Theta$ remains when the first-layer weights connected to the unimportant predictors are set to zero.
These properties of $f^{\ast}_{\btheta}$ and $\Theta$ are summarized in Lemmas 6-7. An additional condition is needed.
\begin{cond}\label{C:Lojasewica}
    There exists a positive constant $\alpha >2$ such that         $\textup{dist}(\btheta,\Theta)^\alpha\leq c_2|U(f_{\btheta})-U(f^{\ast}_{\btheta})|$,
    where $c_2$ is a positive constant, and $\textup{dist}(\btheta,\Theta)\coloneqq \min_{\bvartheta\in\Theta}\|\btheta-\bvartheta\|_{2}$.
\end{cond}
This is a technical condition and has been assumed in the literature. A fixed network with analytic activation functions, including both the classic (linear, ranh, Sigmoid) and more recent ReLU-type (GeLU, ELU, PeLU) functions, satisfies this condition, which can be established through \L ojasewica's inequality. Note that $\alpha$ can be upper-bounded by some positive constants in a polynomial function class, as proved by \cite{bolte2010characterizations} and \cite{Lojasiewicz2012}. Given the close relationship between the polynomial function class and the network class \citep{li2020powernet, shen2024nonparametric}, this condition is sensible.

\begin{theorem}\label{thm:conv-p-nmrc}
    Suppose that Conditions \ref{C:mathcal-F}-\ref{C:indicator-bound} hold. If $\calF_{\btheta_n}$ satisfies (\ref{eq:depth-width-NN-high-p}) and its weights are bounded by a constant $0<B_2<\infty$, then $\hat{f}_{\btheta_{\lambda_n}}$ defined in (\ref{est:P-NMRC}) satisfies: 
    \begin{align}\label{eq:conv-p-nmrc-1}
         &\nonumber\bbE[U(f^{\ast})-U(\hat{f}_{\btheta_{\lambda_n}})]\\ \precsim &\lambda_n p(N\log N)^{1/2} + (N M)^{-2 \beta_1 / p_s} + p(N\log N) (M\log M) \sqrt{\frac{\log(pNM)\log n}{n}}.
    \end{align} 
    When $p\ll n^{1/2}$ and under suitable choices of $M$, $N$ and $\lambda_n$, (\ref{eq:conv-p-nmrc-1}) converges to zero  $n\to \infty$.
\end{theorem}
The bounded weights assumption has also been commonly made in deep network studies \citep{dinh2020consistent,Fan2023Factor,shen2023complexity}. While there is an additional error term compared to the upper bound of the non-penalized estimator $\hat{f}_{\btheta}$ given in Theorem \ref{thm:convergence-rate-high}, the same convergence rate can be achieved with appropriate choices of $M$, $N$, and $\lambda_n$. Let $\bmu(\hat{\btheta}_{\lambda_n})$ and $\bnu(\hat{\btheta}_{\lambda_n})$ denote the components of the first-layer weight matrix that connect to $\bX_{\calJ}$ and $\bX_{\calJ^{c}}$, respectively. The following result can be established.

\begin{theorem}\label{thm:conv-parameter-level}
Suppose that Conditions \ref{C:mathcal-F}-\ref{C:Lojasewica} hold. If $\calF_{\btheta}$ satisfies (\ref{eq:depth-width-NN-high-p}) and its weights are bounded by a constant $0<B_2<\infty$, then when $(NM)^{-2\beta_1/p_s}\to0$, 
    \begin{align}
         \bbE[\textup{dist}(\hat{\btheta}_{\lambda_n},\Theta)^\alpha] \precsim &p(N\log N) (M\log M) \sqrt{\frac{\log(pNM)\log n}{n}} +(\lambda_n \sqrt{p})^{\frac{\alpha}{\alpha-1}}, \label{eq:conv-theta}\\
         \bbE\|\bm{\nu}(\hat{\btheta}_{\lambda_n})\|_2\precsim& \lambda_n^{-1}p(N\log N)(M\log M)\sqrt{\frac{\log(pNM)\log n}{n}}+\sqrt{p}\bbE[\textup{dist}(\hat{\btheta}_{\lambda_n},\Theta)],\label{eq:conv-nu-theta} 
    \end{align}
where $\alpha>2$ is a positive constant. If $p\ll n^{\frac{2\beta_1}{4\beta_1+\alpha(2\beta_1+p_s)}}$ and the smooth index $\beta_1>\frac{\alpha p_s}{2\alpha-4}$, under suitable choices of $M$, $N$, and $\lambda_n$,  (\ref{eq:conv-theta})-(\ref{eq:conv-nu-theta}) converge to zero as $n\to\infty$.
\end{theorem}
With this result, the first-layer weights connecting to the important predictors are bounded away from zero, and those connecting to the unimportant variables converge to zero. As such, variable selection can be achieved based on $\hat{f}_{\btheta_{\lambda_n}}$. In what follows, we examine some special cases of Theorems \ref{thm:conv-p-nmrc} and \ref{thm:conv-parameter-level}.

\begin{corollary}\label{cor:variable-selection-special-cases}
    Suppose that Conditions \ref{C:mathcal-F}-\ref{C:Lojasewica} hold, $\calF_{\btheta}$ satisfies (\ref{eq:depth-width-NN-high-p}), and its weights are bounded by a constant $0<B_2<\infty$. Then,  (\ref{eq:conv-p-nmrc-1})-(\ref{eq:conv-nu-theta}) imply:
    \begin{itemize}
        \item[(i)] if  $p=O(\log^{a_1} n)$ with $0\leq a_1<\infty$,  we set $M=O(n^{\frac{p_s}{2(2\beta_1+p_s)}})$, $N=O(1)$ and $\lambda_n=O(n^{-\frac{\beta_1}{2(2\beta_1+p_s)}})$, then  $\hat{f}_{\btheta_{\lambda_n}}$ and $\hat{\btheta}_{\lambda_n}$ satisfy:  
        \begin{align*}
             &\bbE[U(f^{\ast})-U(\hat{f}_{\btheta_{\lambda_n}})] \precsim n^{-\frac{\beta_1}{2(2\beta_1+p_s)}} \log^{a_1+3} n, \\ 
            &\bbE[\textup{dist}(\hat{\btheta}_{\lambda_n},\Theta)] \precsim n^{-\frac{\beta_1}{2 (\alpha-1)(2\beta_1+p_s)}} \log^{\frac{a_1+3}{\alpha}} n, \\ &\bbE\|\bm{\nu}(\hat{\btheta}_{\lambda_n})\|_2\precsim n^{-\frac{\beta_1}{2 (\alpha-1)(2\beta_1+p_s)}} \log^{\frac{(\alpha+1)a_1+3}{\alpha}} n,
        \end{align*}
        which converge to zero as $n\to\infty$;
        \item[(ii)] if $p=O(n^{a_2})$ with $0<a_2<\frac{2\beta_1}{4\beta_1+\alpha(2\beta_1+p_s)}$ and  $\beta_1>\frac{\alpha p_s}{2\alpha-4}$, we set $M=O(n^{\frac{p_s}{2\beta_1+p_s}(\frac{1}{2}-a_2)})$, $N=O(1)$, and $\lambda_n=O(n^{-\frac{2\alpha\beta_1}{4\beta_1+\alpha(2\beta_1+p_s)}})$, then $\hat{f}_{\btheta_{\lambda_n}}$ and $\hat{\btheta}_{\lambda_n}$ satisfy: 
     \begin{align*}
            &\bbE[U(f^{\ast})-U(\hat{f}_{\btheta_{\lambda_n}})]\precsim n^{-\frac{2\beta_1}{2\beta_1+p_s}(\frac{1}{2}-a_2)}\log^3 n, \\ 
            &\bbE[\textup{dist}(\hat{\btheta}_{\lambda_n},\Theta)]\precsim n^{-\frac{2\beta_1}{\alpha(2\beta_1+p_s)}(\frac{1}{2}-a_2)}\log^3 n,\\ 
            &\bbE\|\bm{\nu}(\hat{\btheta}_{\lambda_n})\|_2\precsim n^{-\frac{\beta_1(1-2 a_2)}{\alpha(2\beta_1+p_s)}+\frac{2\alpha\beta_1}{4\beta_1+\alpha(2\beta_1+p_s)}}\log^3 n,
        \end{align*}
        which converge to zero as $n\to\infty$.
\end{itemize}
\end{corollary}
When $p$ is fixed, the convergence rates of $\bbE[\textup{dist}(\hat{\btheta}_{\lambda_n},\Theta)]$  and $\bbE\|\bm{\nu}(\hat{\btheta}_{\lambda_n})\|$ are the same. When $p$ increases with $n$, $\bbE[\textup{dist}(\hat{\btheta}_{\lambda_n},\Theta)]$ converges to zero faster than $\bbE\|\bm{\nu}(\hat{\btheta}_{\lambda_n})\|$. The theoretical results for the penalized DReLU estimator $\hat{f}_{\btheta_{\omega_n,\lambda_n}}$ are analogous, can be derived following similar steps, and are omitted here.

\section{Accommodation of censored survival data}\label{sec:extension-censored}

\subsection{Methods}

Let $T$ be the event time and $C$ be the censoring time. 
The observations $\{(\bX_i,Y_i,\Delta_i)\}_{i=1}^n$ are i.i.d. copies of $(\bX,Y,\Delta)$, where $Y=\min(T,C)$ and $\Delta=I(T\leq C)$. Here the notations have similar implications as above. 
Consider the model: 
\begin{align}\label{Model-censor}
    T=D\circ F(f^{\ast}(\bX),\epsilon).
\end{align}
We propose objective function and estimator:
\begin{align}\label{obj:C-S-NMRC}
    U^{c}_{n}(f)=&\frac{1}{n(n-1)}\sum_{i\neq j}\Delta_j I(Y_i>Y_j)I(f(\bX_i)>f(\bX_j)),
\end{align}
\begin{align}\label{est:C-S-NMRC}
    \hat{f}^{c}_{\btheta}\in\argmax_{f_{\btheta}\in\calF_{\btheta}}U^{c}_{n}(f_{\btheta}).
\end{align}
The parameter of the estimated network $\hat{f}^{c}_{\btheta}$ is denoted by $\hat{\btheta}^{c}$. This estimation has been motivated by the developments above and \cite{ He2023JASA} and others under the regression framework. Further, we consider the DReLU objective function:
\begin{align*}
     U^{c}_{n,\omega_n}(f)=\frac{1}{n(n-1)}\sum_{i\neq j}\Delta_j I(Y_i>Y_j)S_{\omega_n}(f(\bX_i)-f(\bX_j)).
\end{align*}
Its maximizer is denoted as $\hat{f}^{c}_{\btheta_{\omega_n}}$. Given the similarity of $\hat{f}^{c}_{\btheta}$ and $\hat{f}^{c}_{\btheta_{\omega_n}}$, in what follows, we focus on the estimate defined in (\ref{est:C-S-NMRC}).

\subsection{Statistical properties}\label{subsec:theory-censor}
We define the population objective function of (\ref{obj:C-S-NMRC}) as: 
\begin{align}\label{obj:C-S-NMRC-pop}
    U^{c}(f)=\bbE[\Delta_2 I(Y_1>Y_2)I(f(\bX_1)>f(\bX_2))].
\end{align}
Additional conditions are assumed.

\begin{cond}\label{C:epsilon-indep}
   $\epsilon$ is independent of $C$ and $\bX$.
\end{cond}

\begin{cond}\label{C:censor-rate}
   The set $\{\bX\in\calX:\Pr(\Delta=1|\bX)>0\}$ has a positive measure.
\end{cond}

\begin{cond}\label{C:indep-censor}
$C$ is independent of $\bX$. 
\end{cond}
Conditions \ref{C:epsilon-indep}-\ref{C:censor-rate} are comparable to their regression counterparts. Condition \ref{C:indep-censor} is imposed for technical convenience and can be relaxed when $F(\cdot,\cdot)$ in model (\ref{Model-censor}) is additive. The following results can be established.

\begin{lemma}\label{lem:identify-Survival}
Under Conditions \ref{C:mathcal-F}-\ref{C:strictly-mono} and \ref{C:epsilon-indep}-\ref{C:indep-censor}, target function $f^{\ast}$ is the unique maximizer of $U^{c}(f)$. 
\end{lemma}

\begin{theorem}\label{lem:convergence-rate-censor}
 Suppose that Conditions \ref{C:mathcal-F}-\ref{C:indep-censor} hold. If $p$ is fixed and $\mathcal{F}_{\btheta}$ satisfies (\ref{eq:depth-width-NN-low-p}), $\hat{f}^{c}_{\btheta}$ defined in (\ref{est:C-S-NMRC}) satisfies:
    \begin{align}\label{eq:conv-rate-censor-low}
       \bbE[U^{c}(f^{\ast})-U^{c}(\hat{f}^{c}_{\btheta})]\precsim(N M)^{-2 \beta / p}+(N\log N) (M\log M) \sqrt{\frac{\log(NM)\log n}{n}}.
   \end{align}
If $p\ll n^{1/2}$ and $\calF_{\btheta}$ satisfies (\ref{eq:depth-width-NN-high-p}), then $\hat{f}^{c}_{\btheta}$ satisfies: 
    \begin{align}\label{eq:conv-rate-censor-high}
       \bbE[U^{c}(f^{\ast})-U^{c}(\hat{f}^{c}_{\btheta})]\precsim(N M)^{-2 \beta_1 / p_s}+p(N\log N) (M\log M) \sqrt{\frac{\log(pNM)\log n}{n}}.
   \end{align}
    Under appropriate choices of $N$ and $M$, (\ref{eq:conv-rate-censor-low}) and (\ref{eq:conv-rate-censor-high}) converge to zero as $n\to\infty$. 
\end{theorem}
The convergence rates established in Theorem \ref{lem:convergence-rate-censor} are the same as those for uncensored data. Censoring plays a role through the upper bound of $\bbE[U^{c}(f^{\ast})-U^{c}(\hat{f}^{c}_{\btheta})]$.

\section{Simulation}\label{sec:simulation}
In simulation, we compare against multiple highly relevant alternatives,  including the deep Least Squares Estimation (NN-LSE) \citep{jiao2023AOS} and various deep robust estimation techniques \citep{shen2021robust}. 
The latter category includes the deep Least Absolute Deviation (NN-LAD) approach which has loss function $L(f_{\btheta})=\frac{1}{n}\sum_{i=1}^{n}|y_i-f_{\btheta}(\bx_i)|$, the deep Cauchy (NN-Cauchy) approach which has loss function $L(f_{\btheta})=\frac{1}{n}\sum_{i=1}^n \log(1+\iota^2 (f_{\btheta}(\bx_i)-y_i)^2)$ for some $\iota>0$, and the deep Huber (NN-Huber) approach which has loss function $L(f_{\btheta})=\frac{1}{n}\sum_{i=1}^n[ \frac{1}{2}(f_{\btheta}(\bx_i)-y_i)^2I(|f_{\btheta}(\bx_i)-y_i|<\iota)+\{\iota|f_{\btheta}(\bx_i)-y_i|-\iota^2/2\}I(|f_{\btheta}(\bx_i)-y_i|>\iota)]$  for any $\iota>0$. 
For the fairness of comparison, the same network architecture is employed. Additionally, we also compare with the ``classic’’ maximum rank correlation (MRC) approach \citep{han1987non}. All the deep network methods are realized using Pytorch and the stochastic gradient descent algorithm Adam.

In all of the following six models, $\bX$ is generated from a multivariate normal distribution $N(\bm{0},\bm{\Sigma})$ with $\bm{\Sigma}=\{0.3^{|i-j|}\}_{i,j=1,\ldots,p}$ and $p\in\{15,200\}$. 
Further, standard normal errors are considered in Models M\ref{M1}, M\ref{M2} and M\ref{M5}, while heavy-tail errors are considered in Models M\ref{M3}, M\ref{M4} and M\ref{M6} with $\epsilon\sim 0.9N(0,1)+0.1\textup{Cauchy}(0,1)$.

\begin{model}\label{M1}
$Y=\bm{X}^{\top}\bm{b}+\sin((\bm{X}^{\top}\bm{b})^2)+\epsilon$, where $\bm{b}=(0.5\cdot\bm{1}_{8},-0.5\cdot\bm{1}_{7},\bm{0}_{p-15})$. This is a single index model with an identity transformation.
\end{model}
\begin{model}\label{M2}  $Y=D(1+\bX^{\top}\bm{b}+\cos((\bm{X}^{\top}\bm{b})^2)+\exp(1-(\bX^{\top}\bm{b})^2)+\epsilon)$,
where $D(u)=uI(u>0)+3uI(u<0)$, $\bm{b}=(0.5\cdot\bm{1}_{8},-0.5\cdot\bm{1}_{7},\bm{0}_{p-15})$.
This is a single index model with a non-identity transformation.
\end{model}
\begin{model}\label{M3}
    $Y=\frac{1}{2}X_1^2 + X_2 X_3 + \frac{1}{3}(X_4-X_5)^2-(X_6^2+1)^{-1}-(X_7^2+1)^{-1}-\sin(X_8) + \sin(X_9 X_{10}) + \exp((X_{11}^{2}+1)^{-1}) + \exp(\frac{1}{2}|X_{12}|) - \exp(\frac{1}{2}X_{13})+ \tanh(X_{14})-\tanh(X_{15}^{-1})+ \epsilon$. This is a nonlinear model.
\end{model}
\begin{model}\label{M4} 
    $Y=\frac{1}{2}(X_1^2+X_2^2)+X_3^2(2+X_4^2)^{-1}-(\frac{1}{2}+X_5)^{-1}+\sin(X_6+X_7)-\sin(X^2_8)+\frac{1}{3}(X^2_9+X^2_{10})+\exp(\frac{1}{2}X_{11})-\exp((1+X^2_{12})^{-1})+\tanh(X_{13})+|X_{14}X_{15}|+\epsilon$. This is a nonlinear model.
\end{model}
\begin{model}\label{M5}
$\log(Y)=\sqrt{\bX^{\top}\bm{b}}+\sin(\bX^{\top}\bm{b})-1+\epsilon/\sqrt{2}$, and $C\sim \mathcal{X}^2(2)$. \\$\bm{b}=(-0.9,0.8,-0.7,0.6,-0.5,0.4,-0.3,0.2,-0.3,0.4,0.5,-0.6,0.7,-0.8,0.9,\bm{0}_{p-15})$. This is a single index accelerated failure time (AFT) model with a log transformation.
\end{model}
\begin{model}\label{M6}
    $Y=\frac{1}{2}(X_1^2+X_2^2+X_3^2)+\sin(X_4+X_5)-\exp(\frac{1}{3}(X_6X_7+X_8))+\exp((X^2_9+1)^{-1})+\exp(\frac{1}{2}X_{10})+\tanh(X_{11}X_{12})-X_{13}(\frac{1}{2}+X^2_{14}+X^2_{15})^{-1}+\epsilon$ and  $C\sim \mathcal{X}^2 (6)$. 
This is a fully nonlinear AFT model.
\end{model}

The censoring rates for M\ref{M5} and M\ref{M6} are about 20$\%$ and 30$\%$, respectively. When $p=200$, group Lasso is applied to the first-layer weights of all the deep network methods, and  Lasso is applied to the ``classic" MRC approach.
For M\ref{M5} and M\ref{M6},  the Kaplan-Meier-based weighted loss function are adopted for the compared methods.  The sizes of the training, validation (used for network parameter tuning), and testing (used for prediction evaluation) data are all 1,000. We simulate 100 replicates for each model.

\begin{table}[h]
    \centering
    \caption{Simulation results. 
    In each cell, mean (sd). }
    \label{tab:sim-1}
     \begin{threeparttable}
    \resizebox{\textwidth}{!}{\begin{minipage}{\textwidth}
    \begin{tabular}{cc ccccc c  ccccccc}
    \hline
    & & \multicolumn{5}{c}{$p=15$}& &\multicolumn{7}{c}{$p=200$} \\ 
    \cline{3-7}\cline{9-15}
      & Method & MSE & LAD & Cauchy & Huber & Rank && MSE & LAD & Cauchy & Huber & Rank & TOP10 & TOP20 \\
     \hline
      \multirow{6}{*}{M\ref{M1}}  & Proposed & {\bf 1.31(0.06)} & {\bf 0.92(0.03)} & {\bf 0.64(0.02)} & {\bf 0.59(0.02)} & {\bf 0.78(0.01)} && {\bf 1.33(0.05)} & {\bf 0.92(0.02)} & {\bf 0.64(0.02)} & {\bf 0.60(0.02)} & {\bf 0.77(0.01)} & {\bf 10.0(0.0)} & {\bf 15.0(0.0)}\\
     & NN-LSE & 1.33(0.06) & {\bf 0.92(0.02)} & {\bf 0.64(0.02)} & {\bf 0.59(0.02)} & {\bf 0.78(0.01)} && 1.37(0.07) & 0.93(0.02) & 0.65(0.02) & 0.61(0.02) & {\bf 0.77(0.01)} & {\bf 10.0(0.0)} & {\bf 15.0(0.0)} \\
     & NN-LAD & 1.34(0.05) & 0.93(0.02) & {\bf 0.64(0.02)} & 0.60(0.02) & {\bf 0.78(0.01)} && 1.43(0.08) & 0.95(0.03) & 0.67(0.03) & 0.63(0.03) & 0.76(0.01) & {\bf 10.0(0.0)} & {\bf 15.0(0.0)}  \\
     & NN-Cauchy & 1.33(0.06) & {\bf 0.92(0.02)} & {\bf 0.64(0.02)} & 0.60(0.02) & {\bf 0.78(0.01)} && 1.42(0.08) & 0.95(0.03) & 0.66(0.03) & 0.63(0.03) & 0.76(0.01) & {\bf 10.0(0.0)} & {\bf 15.0(0.0)}\\
     & NN-Huber & 1.33(0.05) & {\bf 0.92(0.02)} & {\bf 0.64(0.02)} & 0.60(0.02) & {\bf 0.78(0.01)} && 1.39(0.08) & 0.94(0.03) & 0.65(0.02) & 0.62(0.03) & 0.76(0.01) & {\bf 10.0(0.0)} & {\bf 15.0(0.0)}  \\
     & MRC & 1.35(0.07) & 0.93(0.03) & 0.65(0.02) & 0.61(0.03) & {\bf 0.78(0.02)} && 1.51(0.19) & 0.98(0.06) & 0.69(0.05) & 0.66(0.07) & 0.75(0.04) & {\bf 10.0(0.0)} & {\bf 15.0(0.0)}   \\
     \hline
     \multirow{6}{*}{M\ref{M2}}  & Proposed & {\bf 4.24(0.32)} & {\bf 1.44(0.04)} & {\bf 0.99(0.02)} & {\bf 1.25(0.05)} & {\bf 0.83(0.01)} && {\bf 4.84(0.31)} & {\bf 1.61(0.05)} & {\bf 1.12(0.04)} & {\bf 1.44(0.06)} & {\bf 0.71(0.04)} & {\bf 10.0(0.0)} & {\bf 15.0(0.0)} \\
     & NN-LSE & 4.57(0.33) & 1.56(0.06) & 1.09(0.04) & 1.38(0.07) & 0.74(0.05) && 4.93(0.33) & 1.65(0.05) &  1.16(0.03) & 1.48(0.06) & 0.68(0.03) &  {\bf 10.0(0.0)} & {\bf 15.0(0.0)}\\
     & NN-LAD & 4.48(0.37) & 1.50(0.04) & 1.04(0.03) & 1.31(0.06) & 0.80(0.02) &&  5.09(0.40) & 1.64(0.05) & 1.14(0.03) & 1.47(0.06) & 0.70(0.02) &  {\bf 10.0(0.0)} & {\bf 15.0(0.0)} \\
     & NN-Cauchy & 4.50(0.43) & 1.49(0.05) & 1.02(0.03) & 1.30(0.07) & 0.81(0.01) && 5.46(0.58) & 1.69(0.08) & 1.18(0.05) & 1.55(0.10) & 0.70(0.03) &  {\bf 10.0(0.0)} & {\bf 15.0(0.0)} \\
     & NN-Huber & 4.43(0.42) & 1.49(0.05) & 1.03(0.03) & 1.31(0.07) & 0.81(0.02) && 4.97(0.38) & 1.63(0.04) & 1.13(0.02) & 1.46(0.06) & 0.70(0.03) &  {\bf 10.0(0.0)} & {\bf 15.0(0.0)} \\
     & MRC & 10.32(0.58) & 2.38(0.06) & 1.60(0.04) & 2.41(0.08) & 0.64(0.02) && 10.90(0.76) & 2.45(0.07) & 1.64(0.05) & 2.50(0.09) & 0.60(0.02) &  {\bf 10.0(0.0)} & {\bf 15.0(0.0)} \\
     \hline
 \multirow{6}{*}{M\ref{M3}} & Proposed & {\bf  2.40(0.15)} & {\bf  1.22(0.04)} & {\bf  0.88(0.03)} & {\bf  0.96(0.04)} & {\bf  0.74(0.01)} && {\bf  3.62(0.20)} & {\bf  1.50(0.04)} & {\bf  1.09(0.04)} & {\bf  1.29(0.05)} & {\bf  0.59(0.03)} & {\bf  8.9(1.0)} & {\bf  10.9(1.5)} \\
     & NN-LSE & 3.81(0.86) & 1.53(0.17) & 1.11(0.12) & 1.33(0.21) & 0.55(0.15) && 5.44(0.87) & 1.82(0.15) & 1.30(0.10) & 1.68(0.19) & 0.31(0.09) & 3.7(2.0) & 4.6(1.8) \\
     & NN-LAD & 2.43(0.11) & 1.23(0.03) & 0.89(0.02) & {\bf  0.96(0.04)} & 0.73(0.02) && 4.77(0.24) & 1.71(0.05) & 1.23(0.04) & 1.55(0.06) & 0.41(0.03) & 5.6(1.6) & 6.6(2.1) \\
     & NN-Cauchy & 2.53(0.14) & 1.26(0.04) & 0.91(0.03) & 0.99(0.04) & 0.72(0.02)  &&  4.85(0.35) & 1.73(0.06) & 1.24(0.04) & 1.57(0.08) & 0.41(0.06) & 6.1(2.4) & 7.4(3.0) \\
     & NN-Huber & 2.41(0.07) & 1.23(0.02) & 0.89(0.01) & {\bf  0.96(0.02)} & {\bf  0.74(0.01)} &&   4.41(0.30) & 1.65(0.05) & 1.20(0.03) & 1.48(0.06) & 0.48(0.04) & 8.3(1.6) & 10.0(2.4) \\
     & MRC & 4.84(0.20) & 1.72(0.04) & 1.24(0.03) & 1.56(0.05) & 0.37(0.03) &&  5.57(0.35) & 1.85(0.06) & 1.33(0.04) & 1.73(0.07) & 0.19(0.08) & 1.8(0.8) & 3.1(1.0)\\
     \hline
 \multirow{6}{*}{M\ref{M4}} & Proposed & {\bf  2.15(0.09)} & {\bf  1.16(0.03)} & {\bf  0.84(0.02)} & {\bf  0.88(0.03)} & {\bf  0.79(0.02)}  && {\bf  2.95(0.17)} & {\bf  1.36(0.04)} & {\bf  0.99(0.03)} & {\bf  1.11(0.05)} & {\bf  0.70(0.03)} & {\bf  9.5(0.6)} & {\bf  11.4(0.6)}\\
     & NN-LSE &  3.83(0.87) & 1.53(0.17) & 1.11(0.12) & 1.33(0.21) & 0.62(0.10) && 5.72(0.95) & 1.86(0.16) & 1.33(0.10) & 1.74(0.20) & 0.31(0.13) & 4.2(1.8) & 4.9(2.2)\\
     & NN-LAD & 2.88(0.11) & 1.21(0.03) & 0.87(0.02) & 0.92(0.03) & 0.77(0.02) && 4.26(0.23) & 1.63(0.05) & 1.18(0.04) & 1.44(0.06) & 0.54(0.03) & 8.6(0.9) & 10.3(1.1)  \\
     & NN-Cauchy & 2.32(0.13) & 1.21(0.03) & 0.88(0.02) & 0.93(0.04) & 0.77(0.02)
  && 4.25(0.17) & 1.63(0.04) & 1.18(0.03) & 1.44(0.05) & 0.54(0.03) & 8.7(0.9) & 10.4(1.2)\\
     & NN-Huber &  2.25(0.15) & 1.19(0.04) & 0.86(0.03) & 0.91(0.04) & 0.78(0.02) && 3.91(0.16) & 1.56(0.04) & 1.13(0.03) & 1.36(0.04) & 0.58(0.03) & 8.7(0.7) & 10.1(1.0) \\
     & MRC &  5.21(0.30) & 1.81(0.05) & 1.29(0.03) & 1.68(0.05) & 0.36(0.03)  && 6.11(0.27) & 1.92(0.05) & 1.36(0.02) & 1.81(0.04) & 0.08(0.07) & 2.3(1.6) & 3.8(1.8)  \\
     \hline
     \multirow{6}{*}{M\ref{M5}}  & Proposed & {\bf 2.54(1.03)} & {\bf 0.90(0.10)} & {\bf 0.57(0.05)} & {\bf 0.70(0.12)} & {\bf 0.73(0.01)} &&  {\bf 3.16(1.49)} & {\bf 1.01(0.12)} & {\bf 0.64(0.06)} & {\bf 0.83(0.15)} & {\bf 0.71(0.01)} & {\bf 10.0(0.0)} & {\bf 15.0(0.0)}\\
     & NN-LSE & 2.80(1.08) & 1.03(0.12) & 0.67(0.07) & 0.80(0.14) & 0.69(0.01) && 4.03(1.14) & 1.39(0.11) & 0.95(0.06) & 1.15(0.13) & 0.52(0.03) & {\bf 10.0(0.0)} & {\bf 15.0(0.0)} \\
     & NN-LAD & 2.77(1.07) & 0.95(0.11) & 0.60(0.06) & 0.75(0.13) & 0.72(0.02) && 4.12(1.43) & 1.20(0.12) & 0.77(0.06) & 1.03(0.14) & 0.59(0.04) & {\bf 10.0(0.0)} & {\bf 15.0(0.0)} \\
     & NN-Cauchy & 2.71(0.99) & 0.92(0.11) & 0.58(0.06) & 0.73(0.13) & 0.72(0.01) && 3.90(1.48) & 1.14(0.13) & 0.72(0.06) & 0.96(0.15) & 0.63(0.03) & {\bf 10.0(0.0)} & {\bf 15.0(0.0)} \\
     & NN-Huber & 2.86(1.40) & 0.99(0.11) & 0.63(0.04) & 0.77(0.14) & 0.71(0.02) && 3.97(1.30) & 1.25(0.11) & 0.81(0.06) & 1.03(0.14) & 0.58(0.04) & {\bf 10.0(0.0)} & {\bf 15.0(0.0)}  \\
     & MRC & 4.38(1.73) & 1.20(0.12) & 0.75(0.05) & 1.02(0.16) & 0.63(0.01) && 4.34(1.63) & 1.23(0.13) & 0.79(0.06) & 1.05(0.16) & 0.59(0.03) & 8.4(2.1) & 10.3(2.4)
     \\
     \hline
     \multirow{6}{*}{M\ref{M6}} & Proposed & {\bf 2.48(0.43)} & {\bf 1.19(0.04)} & {\bf 0.85(0.04)} & {\bf 0.91(0.05)} & {\bf 0.73(0.02)} && {\bf 3.90(0.53)} & {\bf 1.48(0.05)} & {\bf 1.06(0.03)} & {\bf 1.26(0.07)} & {\bf 0.67(0.01)} & {\bf 8.0(0.6)} & {\bf 8.3(1.4)} \\
     & NN-LSE & 3.81(0.46) & 1.50(0.08) & 1.81(0.06) & 1.28(0.09) & 0.69(0.02) && 6.58(0.61) & 1.94(0.08) & 1.36(0.05) & 1.84(0.11) & 0.55(0.02) & 1.6(1.2) & 2.8(1.4) \\
     & NN-LAD & 3.09(0.48) & 1.34(0.07) & 0.97(0.05) & 1.09(0.08) & 0.72(0.02) && 5.29(0.67) & 1.72(0.09) & 1.26(0.06) & 1.63(0.12) & 0.61(0.01) & 5.5(0.8) & 6.5(1.2)  \\
     & NN-Cauchy & 2.96(0.46) & 1.31(0.06) & 0.94(0.04) & 1.05(0.07) & 0.72(0.02) && 5.14(0.70) & 1.71(0.10) & 1.24(0.06) & 1.61(0.13) & 0.61(0.01) & 5.1(0.7) & 5.8(1.0) \\
     & NN-Huber &  2.97(0.48) & 1.32(0.07) & 0.95(0.05) & 1.06(0.08) & 0.72(0.01) && 5.20(0.60) & 1.70(0.08) & 1.24(0.05) & 1.54(0.10) & 0.62(0.01) & 5.8(0.7) & 6.4(1.2)  \\
     & MRC & 4.56(0.82) & 1.59(0.09) & 1.13(0.06) & 1.39(0.11) & 0.64(0.01) && 4.82(0.54) & 1.63(0.08) & 1.15(0.05) & 1.45(0.10) & 0.63(0.01) & 6.0(0.8) & 6.8(1.1)  \\
     \hline
    \end{tabular}
    \end{minipage}}
    \end{threeparttable}
\end{table}

For evaluating prediction performance, we consider Mean Square Error (MSE), Least Absolute Deviation (LAD), Cauchy loss, and Huber loss, all of which have been adopted in the literature. As the rank-based approaches generate ranks as opposed to exact predictive values, we further apply a simple linear transformation step. Additionally, the Spearman correlation coefficient is calculated to describe rank correlation. For survival data, rank correlation is evaluated using the Concordance Index (C-idx).
Both the Spearman correlation coefficient and C-index are referred to as ``Rank" in Table \ref{tab:sim-1}.
As in published studies, variable selection is evaluated by TOP10 (TOP20) when $p=200$, which is the number of true positives among the top 10(20) identified predictors. 
Partial results are summarized in Table \ref{tab:sim-1}. Additional results are provided in supplementary materials, including tuning parameter selections and results under settings with other dimensions and error distributions. The observations made in supplementary materials are similar to those in Table \ref{tab:sim-1}. 

Table \ref{tab:sim-1} shows that when data is generated from a simple model, the proposed method performs comparably to the others. However, as the data generation mechanism becomes more complex, it becomes more advantageous, with smaller prediction errors, larger Spearman correlation coefficients, and higher identification accuracy. It demonstrates significant improvements compared to NN-LSE in Models \ref{M2}-\ref{M6}, highlighting the merit of robustness. Additionally, when $D\circ F$ is not an identity function, the proposed method outperforms the other robust estimations, demonstrating its broader applicability. Compared to the ``classic" MRC, it exhibits slightly better performance in M\ref{M1} and substantial advantages across all the criteria when data is generated from more complex models (M\ref{M2}-M\ref{M6}). This re-establishes the necessity of adopting deep learning to capture complex nonlinear relationships. Furthermore, Table \ref{tab:sim-1} shows that when $f^{\ast}$ is a single index function (M\ref{M1}, M\ref{M2}, M\ref{M6}), almost all the methods can correctly identify the important variables. In the other scenarios, however, the proposed method yields more satisfactory variable selection results.

\section{Data Analysis}\label{sec:real-data}

We analyze five datasets: the Energy Prediction data \citep{Candanedo2017DataDP}, the Communities and Crime data \citep{Redmond2002ADS}, the Molecular Taxonomy of Breast Cancer International Consortium (METABRIC) data \citep{pereira2016somatic}, the South German Credit data \citep{misc_south_german_credit_573}, and the Wiki4HE data \citep{misc_wiki4he_334}. Among them, the Energy Prediction and Communities and Crime data have continuous responses without censoring, the METABRIC data has a censored survival response, and the South German Credit and Wiki4HE data have binary responses. For data with continuous responses without and with censoring, we apply NN-LSE, NN-LAD, NN-Cauchy, NN-Huber, and MRC for comparison. For data with binary responses, we compare with the deep Logistic regression (NN-Logistic), parametric logistic regression (Logistic), Support Vector Machine (SVM), and MRC. Estimation is conducted using the whole data. Prediction evaluation is based 50 random splits, and under each split, the training, validation, and testing data have a ratio of 7:1:2. The implementation details for the proposed method are summarized in Section 9 in  Appendix.

\noindent{\bf The Energy Prediction Data}
This dataset has energy usage as the response variable. It contains 19,735 samples and 26 predictors, which cover environmental conditions, sourced from a Zigbee wireless sensor network, and external weather data obtained from the nearest airport weather stations. As presented in Table A.13, the proposed method identifies six important variables, and humidity in the teenager's room is identified as the most important, according to the $L_2$ norms of the first-layer weights.

\noindent{\bf  The Communities and Crime Data}
This dataset combines socio-economic data from the 1990 US Census, law enforcement data from the 1990 US LEMAS survey, and crime data from the 1995 FBI UCR. The response variable is violent crime rate. There are 120 predictors describing community characteristics and law enforcement attributes. Data is available for 1,994 samples. Imputation is applied to accommodate missingness. Five variables are identified as important (Table A.13).

\noindent{\bf The METABRIC Data}
The original data contains the measurements of 20,000 gene expressions for 1,973 samples. To improve reliability, as in the literature, we focus on the 482 genes in the following relevant KEGG pathways: hsa05204 (Chemical carcinogenesis - DNA adducts), hsa05205 (Proteoglycans in cancer), hsa05208 (Chemical carcinogenesis - reactive oxygen species), and hsa05224 (Breast cancer). We additionally consider 6 clinical variables, which leads to a total of 488 predictors. The response variable is overall survival, which is subject to right censoring. Three variables are identified as important.

\noindent{\bf The South German Credit Data} 
This is an improved and expanded version of the original Statlog German Credit dataset. Data is available for 1,000 credit instances, and the binary response variable indicates whether credit is good or bad. There are 20 predictors, covering the purpose of the credit, the type of housing occupied by the debtor,  the duration of the credit in months, and others. Dummy variables are created for the categorical variables, leading to a total of 37 input variables. Six variables are identified as important. 

\noindent{\bf The Wiki4HE Data}
This data is generated from a survey conducted on the faculty members of two Spanish universities, focusing on the pedagogical applications of Wikipedia. The sample size is 909. The response variable is the utilization of Wikipedia for teaching purposes, coded as binary. The predictors are captured on a Likert scale, ranging from 1 (strongly disagree / never) to 5 (strongly agree / always). There are a total of 50 input variables. Two variables are identified as important. 

The alternative approaches are found to have different identification results. To enhance comparability, in Table A.14 in supplementary materials, we compare the top ten variables identified by the different methods. The prediction results are summarized in Table \ref{tab:real-1}, which shows the competitive performance of the proposed method. 

\begin{table}[h]
    \centering
    \caption{Data analysis: prediction performance. In each cell, mean (sd). }
    \label{tab:real-1}
     \begin{threeparttable}
    \resizebox{\textwidth}{!}{\begin{minipage}{\textwidth}
    \begin{tabular}{c c ccc cccc}
    \hline
   \multicolumn{2}{c}{}  & Proposed$(\lambda_1=0)$ & Proposed$(\lambda_1\neq0)$ & NN-LSE & NN-LAD & NN-Cauchy & NN-Huber & MRC\\
   \hline
    Energy Prediction  & MSE & 0.90(0.03) & 0.90(0.03) & {\bf 0.87(0.03)} & 0.91(0.03) & 0.87(0.03) & 0.87(0.04) & 1.05(0.03)  \\
   & LAD & 0.42(0.06) & {\bf 0.41(0.07)} &  0.51(0.02) & 0.41(0.06) & {\bf 0.41(0.05)} &  0.44(0.09) & 0.54(0.06) \\
   & Cauchy & 0.23(0.04) & {\bf 0.23(0.05)} & 0.29(0.01) & 0.23(0.06) & 0.23(0.04) & 0.24(0.06) & 0.30(0.05)\\
   & Huber & 0.29(0.08) & 0.28(0.08) & 0.29(0.01) & 0.28(0.08) & 0.27(0.07) & {\bf 0.26(0.07)} & 0.31(0.05) \\
   & Spearman & 0.69(0.01) & {\bf 0.70(0.09)} & 0.58(0.02) & 0.69(0.07) & 0.67(0.02) & 0.64(0.02) & 0.33(0.04)\\
   \hline
  Communities Crime  & MSE & {\bf 0.19(0.02)} & {\bf 0.19(0.02)} & {\bf 0.19(0.02)} & 0.20(0.02) & {\bf 0.19(0.02)} & {\bf 0.19(0.02)} & 0.31(0.05) \\
   & LAD & 0.91(0.05) & {\bf 0.90(0.05)} & 0.97(0.03) & 0.92(0.03) & 0.96(0.03) & 0.97(0.03) & 1.19(0.09) \\
   & Cauchy & {\bf 0.18(0.02)} & {\bf 0.18(0.02)} & {\bf 0.18(0.01)} & {\bf 0.18(0.02)} & {\bf 0.18(0.01)} & {\bf 0.18(0.01)} & 0.29(0.04) \\
   & Huber & 0.10(0.01) & {\bf 0.09(0.01)} & 0.10(0.07) & 0.10(0.01) & 0.10(0.08) & 0.10(0.08) & 0.16(0.03) \\
   & Spearman & {\bf 0.83(0.08)} & {\bf 0.83(0.07)} & 0.82(0.08) & 0.82(0.08) & 0.82(0.01) & 0.82(0.09) & 0.70(0.07) \\
    \hline
    METABRIC & MSE & 0.57(0.15) & 0.52(0.17) & 0.64(0.43) & 0.52(0.27) & 0.53(0.38) & {\bf 0.49(0.27)} & 0.62(0.18) \\
   & LAD & 0.49(0.16) & {\bf 0.44(0.13)} & 0.51(0.27) & 0.45(0.20) & 0.45(0.18) & 0.45(0.20) & 0.49(0.18) \\
   & Cauchy & 0.31(0.11) & {\bf 0.27(0.08)} & 0.34(0.21) & 0.28(0.14) & 0.28(0.12) & 0.28(0.14) & 0.33(0.12) \\
   & Huber & 0.26(0.07) & {\bf 0.22(0.10)} & 0.29(0.19) & 0.23(0.12) & 0.23(0.09) & 0.23(0.12) & 0.28(0.09) \\
   & C-idx & 0.63(0.06) & {\bf 0.64(0.03)} & 0.60(0.03) & 0.62(0.03) & 0.62(0.03) & 0.62(0.03) & 0.62(0.03)   \\
   \hline
    \multicolumn{2}{c}{ } & Proposed$(\lambda_1=0)$ & Proposed$(\lambda_1\neq0)$ & NN-Logistic & Logistic & SVM & MRC & -\\
    \hline
     South German Credit & Classification Error & {\bf0.22} & {\bf 0.22} & {\bf0.22} & 0.25 & 0.24 & 0.24 & -\\
     \hline
Wiki4HE & Classification Error & {\bf0.10 }& {\bf 0.10} & 0.12 & 0.12 & 0.12 & 0.11 & -\\
    \hline
    \end{tabular}
    \end{minipage}}
    \end{threeparttable}
\end{table}

\section{Discussion}
A new model and corresponding estimation, which have the unique and much-desired robustness properties, have been developed. The proposed approach can coherently accommodate multiple data types under one umbrella – this property is not broadly shared. As demonstrated in the numerical studies, it can be significantly advantageous under certain data settings and provide a useful alternative to the existing literature. The establishment of the theoretical properties can not only provide a solid ground for the proposed approach but also have independent value, given the unique formulation/structure of the proposed approach. In regression, the MRC technique has been extended to many data scenarios. It can be of interest to conduct parallel research under the DNN framework.

\section*{Supplementary materials}
The supplementary materials contain technical proofs of the theoretical results in Sections \ref{sec:theorey}-\ref{sec:extension-censored}, and additional  results for numerical experiments in Sections \ref{sec:simulation}-\ref{sec:real-data}.

 \section*{Acknowledgements}
 This study is supported by NIH CA204120, CA276790, and HL161691.

\bibliography{ref}

\end{document}